\documentclass[12pt,aps,apsmath]{jpconf}

\usepackage{psfig}
\usepackage{graphicx}

\newcommand{\be}{\begin{eqnarray}}
\newcommand{\ee}{\end{eqnarray}}
\newcommand{\nn}{\nonumber}
\newcommand{\bea}{\begin{eqnarray}}
\newcommand{\eea}{\end{eqnarray}}

\begin{document}

\title{\bf Probing Color Response - Wakes in a Color Plasma}

\author{J\"org Ruppert \dag\ 
\footnote[1]{email:ruppert@phy.duke.edu} }

\address{\dag\ \ddag\  Department of Physics, Duke University,
 \\ Science Drive, Box 90305, Durham 27707-0305, NC, USA}

\begin{abstract}
The wake induced in a hot QCD medium by a high momentum 
parton (jet precursor) is calculated in the framework of linear response
theory. Two different scenarios are discussed: a weakly coupled quark gluon
plasma (pQGP) as described by hard-thermal loop (HTL) perturbation theory and 
a strongly cupled QGP (sQGP) with the properties of a quantum liquid.
In the latter case the wake could exhibit a pronounced Mach cone structure.
This physical mechanism could be important for the understanding of 
preliminary data from the PHENIX and STAR experiments at RHIC on
the angular distribution of low-pt secondaries stemming
from the away-side jet which indicate maxima at $\Delta\phi=\pi \pm 1.1$.
\end{abstract}

\section{Introduction}

QCD Jet quenching in relativistic heavy ion collisions 
has been proposed as an important possible signal of 
the creation of a quark-gluon plasma \cite{Gyulassy:1990ye,Wang:1991vs}.
It is extensively studied at RHIC. 

The radiative energy loss of the leading parton due to the emission of a secondary partonic shower has been the main emphasis in the theoretical studies so 
far. For an overview of the quantum field theoretical calculations of the radiative energy loss and the experimental implications, see e.~g.~\cite{Accardi:2003gp,Jacobs:2004qv,Kovner:2003zj}.

The study \cite{Wake} focussed on a different aspect of the QCD jet physics
in medium, namely the properties of the wake of current density, charge density and (chromo-)electirc and magnetic field configurations that is induced
by a charged particle traveling through a QCD medium with high momentum.
The present contribution relies on that work \cite{Wake}.

The coherent behavior of the plasma as a reaction of a charged particle traveling through it is investigated by calculating the plasma's response to 
that external current. The calculation is restricted to only one weak, external current which points in a fixed direction in color group space. Methods of 
linear response theory are applied. The medium is assumed to be isotropic and homogenous. In such a framework quantum and non-ablian effects are included indirectly via the dielectric functions, $\epsilon_L$ and $\epsilon_T$. Linear
response theory implies that the dielectric functions are not in turn modified by the effects of the external current.

Two qualitatively different scenerarios are investigated: In the first one, one
assume that the plasmon is in a high temperature regime. There the gluon self-energies can be described using a high temperature expansion \cite{Weldon,Klimov} $T\gg \omega,k$. In the second scenario we study what are the significant physical effects
if the plasma is strongly coupled (sQGP). In that scenario the plasma might
be described best in terms of a quantum liquid. 

It will be shown that the wake of a jet in such a quantum liquid 
can exhibit a characteristic cone-structure
in the charge- and current-densities under certain conditions for the plasmon's dispersion relation and
assuming a `supersonical' velocity of the jet . Since it can be expected that 
the phenomenon of these Mach-cones leads to correlations in the directed emission of secondary partons from the plasmon. Observable consequences for 
particle correlations of this effect will be discussed in detail later in the present contribution.

The idea of determining the sound velocity of plasmon excitations of the expanding plasma from the emission pattern of the plasma particles traveling at an angle with respect to the jet axis has been discussed in \cite{Stoecker, Stoecker2}. 
Our consideration of Mach-cones have also been motivated in parts
by earlier studies of wakes induced by fast electrically charged projectiles in the electron plasma of metallic targets \cite{Schaefer1,Schaefer2}, as well as by recent work exploring the induction of a conical flow by a jet in a sQGP \cite{Shuryak}. This latter study invokes a hydrodynamical description of the energy deposited by a quenched jet in the medium and emphasizes the emergence of a Mach cone.

This proceeding contribution is organized as follows: first, in section \ref{JRplasma}, 
I give a short reminder of some general properties of plasma physics and 
derive the equation for the induced charge- and current-density in the considered situation. In section \ref{JRpQGP} I discuss the wake as obtained in a pQGP. In section \ref{JRsQGP} the wake obtained in a strongly coupled QGP is discussed. Possible 
observable consequences for particle correclations are studied in section \ref{JRobservable}. In section \ref{JRconclusions} I end with conclusions and a short acknowledgment. 

The reader solely interested in the {\it results} related to 
a Mach cone structure might actually skip the next two sections and continue at 
section \ref{JRsQGP}.

\section{Plasma physics}
\label{JRplasma}
The formalism for linear response theory of a plasma can straightforwardly be
generalized from electromagnetic plasma physics (see e.~g.~\cite{Ichimaru}). 

A dielectric medium can be characterized by the components the dielectric tensor $\epsilon_{ij}$, which can be decomposed into  $\epsilon_L$ and $\epsilon_T$
for an isotropic and homogenous medium via the longitudinal and transverse
orthonormal projectors ${\cal P}_{L,ij}=k_i k_j/ k^2$ and ${\cal P}_T=1-{\cal P}_L$ with respect to the momentum vector $\vec{k}$.
Invoking a quantum field theoretical description one can relate these 
dielectic functions to the self-energies $\Pi_L$ and $\Pi_T$ of the in-medium
gluon \cite{LeBellac}
\bea
\epsilon_L(\omega,k)=1-\frac{\Pi_L(\omega,k)}{K^2}, \\
\epsilon_T(\omega,k)=1-\frac{\Pi_T(\omega,k)}{\omega^2}\,,
\eea
where $K^2=\omega^2-k^2$.

The knowledge of these dielectric tensors allows to relate the external 
current $\vec{j}_{\rm ext}$ to the total chromoelectric field $\vec{E}^a_{\rm tot}$ in the QCD plasma via:
 \begin{eqnarray} \label{JREtotTOJ}
 \left[\epsilon_L {\cal P}_{L} + \left(\epsilon_T -\frac{k^2}{\omega^2}\right) {\cal P}_{T}\right] \vec{E}^a_{\rm tot}(\omega,\vec{k})=\frac{4 \pi}{i \omega} \vec{j}^a_{\rm ext} (\omega,\vec{k}) .
 \end{eqnarray}

The color charge induced by the external charge distribution is
\bea \label{JRcharge}
\rho_{\rm ind}=\left(\frac{1}{\epsilon_L}-1\right)\rho_{\rm ext}.
\eea

The induced color charge density can also be calculated from the induced scalar potential via a Poisson equation:
\bea
\Phi_{\rm ind}= \frac{4\pi}{k^2} \rho_{\rm ind} ,
\eea
if one works in a gauge where the vector potential is transverse to the momentum \cite{Neufeld,Ichimaru}.

Since one can relate the total chromo-electric field to the induced charge in linear response theory by 
\bea
\vec{j}^a_{\rm ind}=i\omega(1-\epsilon)\vec{E}^a_{\rm tot}/(4\pi) ,
\eea
a direct relation between the external and the induced current can be derived using Eqn. (\ref{JREtotTOJ}):
\bea \label{JRcurrent}
\vec{j}^a_{\rm ind}=\left[\left(\frac{1}{\epsilon_L}-1\right){\cal P}_L + \frac{1-\epsilon_T}{\epsilon_T-\frac{k^2}{\omega^2}} {\cal P}_T\right]\vec{j}^a_{\rm ext}.
\eea

The induced charge and the induced current obey a continuity equation:
\bea \label{JRcontinuity}
i\vec{k}\cdot\vec{j}_{\rm ind}-i\omega \rho_{\rm ind}=0.
\eea

For the following it is helpful to specify the current and charge densities associated with a color charge as Fourier transform of a 
point charge moving along a straight-line trajectory 
with constant velocity $\vec v$:
\begin{eqnarray}
\vec{j}^a_{\rm ext}&=&2\pi q^a \vec{v} \delta(\omega-\vec{v} \cdot \vec{k}),\\
\vec{\rho}^a_{\rm ext}&=&2\pi q^a \delta(\omega-\vec{v} \cdot \vec{k})
\end{eqnarray}
 where $q^a$ is its color charge defined by $q^a q^a = C \alpha_s$, with the strong coupling constant $\alpha_s=g^2/4\pi$ and the quadratic Casimir invariant $C$ (which is either $C_F=4/3$ for quarks or $C_A=3$ for a gluon). 
In this model description changes of the color charge due to interactins while propagating through the medium are disregarded by fixing the charge's orientation in color space \cite{Weldon,Thoma}
The induced charge-  $\rho_{\rm ind}$ for this situation reads in cylindrical coordinates:
\bea \label{JRcharge2}
\rho_{\rm v, ind}(\rho, z, t)=\frac{m_g^3}{(2 \pi)^2 v} q^a \int_0^\infty d\kappa' \kappa' J_0(\kappa' \rho m_g) \int^{\infty}_{-\infty} d\omega' \,{\rm exp} \left[i\omega'\left(\frac{z}{v}-t\right)m_g\right] \left(\frac{1}{\epsilon_{L}}-1 \right),
\eea
 where $k=\sqrt{\kappa^2+\omega^2/v^2}$, $\kappa=\kappa' m_g$ and $\omega=\omega' m_g$, showing that the induced charge density $\rho_{\rm v, ind}$ is proportional to $m_g^3$. The cylindrical symmetry around the jet axis restricts the form of the current density vector $\vec{j}_{\rm ind}$. It has only non-vanishing components parallel to the beam axis, $\vec{j}_{{\rm v}, {\rm ind}}$, and radially perpendicular to it, $\vec{j}_{\rho, {\rm ind}}$:
\bea \label{JRcurrent2}
\vec{j}_{\rm v, ind}(\rho, z,t) &=&
\frac{m_g^3}{(2 \pi)^2 v^2} q^a  \int_0^\infty d\kappa' \kappa' J_0(\kappa' \rho m_g) \times \nn \\ && \int^{\infty}_{-\infty} d\omega' {\rm exp} \left[i\omega'\left(\frac{z}{v}-t\right)m_g\right] \left[\left(\frac{1}{\epsilon_{L}}-1\right) \frac{\omega^2}{k^2} + \frac{1-\epsilon_{T}}{\epsilon_{T}-\frac{k^2}{\omega^2}}\left(v^2-\frac{\omega^2}{k^2}\right) \right], 
\nn \\
\vec{j}_{\rho, ind}(\rho, z ,t) &=& 
\frac{i m_g^3}{(2 \pi)^2 v} q^a \int_0^\infty d\kappa' \kappa' J_1(\kappa' \rho m_g) \times \nn \\ &&\int^{\infty}_{-\infty} d\omega' {\rm exp} \left[i\omega'\left(\frac{z}{v}-t\right)m_g\right] \frac{\omega \kappa}{k^2}\left[\left(\frac{1}{\epsilon_{L}}-1\right)-\left(\frac{1-\epsilon_{T}}{\epsilon_{T}-\frac{k^2}{\omega^2}}\right)\right] .
\eea
Again, the components of the current density are proportional to $m_g^3$.


In order to deepen the understanding of the results obtained in the different scenarios it is helpful 
to discuss the corresponding plasmon dispersion relations for a given dielectric function. In order to determine non-trivial solutions of
Equation (\ref{JREtotTOJ}), the following determinant has to vanish:
\bea
{\rm det}\left|\epsilon_L {\cal P}_{L} + (\epsilon_T -\frac{k^2}{\omega^2}) {\cal P}_{T}\right|=0 .
\eea
This equation governs the dispersion relation for the waves in the medium. Since it can be diagonalized into purely longitudinal and transverse parts, dispersion relations for the longitudinal and transverse dieelectric functions can be derived \cite{Ichimaru}:
\bea \label{JRDispersion}
\epsilon_L&=&0, \\
\epsilon_T&=&(k/\omega)^2.
\eea
These equations determine the longitudinal and transverse plasma modes. The longitudinal equation is also the dispersion relation for density-fluctuations in the plasma, namely space-charge fields which could be spontaneously excited in the plasma without an application of external disturbances \cite{Ichimaru}.

\section{Charge wake in a QGP in the high temperature approximation}
\label{JRpQGP}

In this section the first scenario is studied in detail. Here the medium is assumed to be in the high
temperature regime where the gluon self-energies are described by the leading order of the high temperature expansion $T\gg \omega, k$ \cite{Klimov, Weldon}. The self-energies derived within the HTL approximation have been shown to be gauge invariant \cite{Pisarski} and the dielectric functions
are therefore also gauge invariant.
The dielectric functions read explicitly \cite{LeBellac}:
\bea
 \epsilon_L &=& 
 1+\frac{2m_g^2}{k^2}
 \left[
 1-\frac{1}{2}x
 \left(
 {\rm ln}\left|\frac{x+1}{x-1}\right|-i\pi \Theta \left(1-x^2\right)
 \right)
 \right], 
 \nn \\
\epsilon_T&=&1-\frac{m_g^2}{\omega^2} 
 \left[ 
 x^2+\frac{x(1-x^2)}{2}  
 \left(
 {\rm ln}\left|\frac{x+1}{x-1}\right|-i\pi 
 \Theta 
  \left(
  1-x^2 
  \right) 
 \right)
 \right],
\label{JReps}
\eea

where $x=\omega/k$. 

\begin{figure}
 \par\resizebox*{!}{0.30\textheight}{\includegraphics{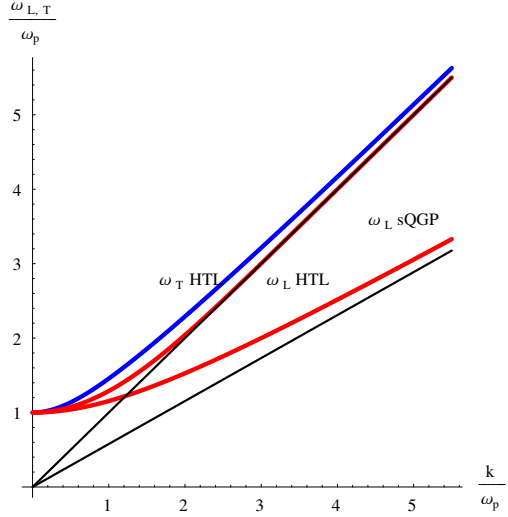}} \par{}
\caption{Longitudinal and transverse plasmon-branches in the $\omega,~k$-plane
for a pQGP (calculation in the HTL approximation) and in the sQGP framework as studied in section \ref{JRsQGP} for (see Eqn. ~(\ref{JRDispersion2}) for $u^2=1/\sqrt{3}$). Note that the pQGP plasmon-branches are only in the time-like region of the plane whereas the
sQGP longitudinal plasmon branch extends above $k>\omega_p/\sqrt{1-u^2}$ into the 
space-like region. The straight lines indicate $\omega=k$ and $\omega=u k$. 
\label{JRfigure0}}
\end{figure}

\begin{figure}
 \par\resizebox*{!}{0.22\textheight}{\includegraphics{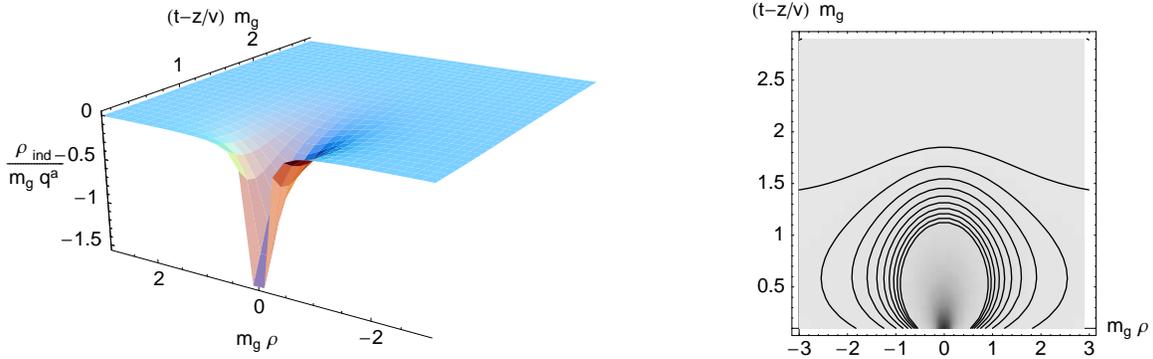}} \par{}
\caption{Spatial distribution of the induced charge density from a jet with fixed color charge $q^a$ in a high temperature plasma where the HTL approximation is applicable. The right plot shows equi-charge lines in the density distribution. The density profile is a cloud traveling with the color charged jet. 
\label{JRfigure1}}
\end{figure}

These dielectric function leads to longitudinal and transverse plasma modes, that are determined by the dispersion relations (\ref{JReps}). They can only appear in the time-like sector of the $\omega,k$ plane \cite{Weldon, LeBellac}, see Fig.~\ref{JRfigure0}. Collective excitations do not contribute to the charge and 
current density profile of the wake. Mach cones do not appear, but the charge carries a screening color cloud 
along with it. Fig.~\ref{JRfigure1} illustrates this physically intuitive result numerically.  
It shows the charge density of a colored parton traveling with $v=0.99 c$ in cylindrical coordinates.
The screening cloud is concentrated in the vincinty of the particle. In the static case where the current
density vector vanishes, the induced charge density can be calculated analytically, it has a Yukawa-like shape $\propto {\rm exp}(-\sqrt{2m_gr})$.

Despite the fact that Mach cones do not appear in the charge density wake,
the particle still suffers energy loss due to elastic collisions in the medium. This effect of energy dissipation has been studied in \cite{Thoma} in detail. The integrand determining the energy loss \cite{Thoma} contributes only where $|x|<1$, and therefore does not get contributions from frequencies 
where collective plasma modes exist. This is consistent with the fact that such modes are not 
excited in the induced charge and current densities.

\section{Charge wake induced in a strongly coupled QGP}
\label{JRsQGP}

Here we assume that the plasma is in a strongly coupled regime having properties of a quantum liquid.
Since there is a lack of theoretical methods for first principle calculations for the dielectric
functions in this regime, we use a simple model calculation in this scenario. Nonetheless, the simplified
model is constructed in such a way that it allows for general conclusion quite independent from 
the exact form of the dielectric functions.

The notion of a sQGP suggest low dissipation at small momenta (`hydro modes') and large dissipation at 
high $k_c$. The existence of a critical momentum $k_c$ is assumed which separates the regimes of collective and single particle modes in the quantum liquid. Below $k_c$ are plasmon excitations and dissipation is
assumed to be neglible for simplicity. Since we are interested in collective effects, the region above $k_c$ is neglected.

To be specific, it is assumed that the longitudinal dielectric function of the strongly coupled plasma in the $k<k_c$ regime is 

\bea \label{JRDispersion2}
\omega_{\rm L}=\sqrt{u^2 k^2 + \omega_p^2}\,\,,
\eea
where $\omega_p$ denotes the plasma frequency and $u$ the speed of plasmon propagation which is assumed 
to be constant. The most important property of this dispersion relation for our purposes is, that
it extends into the space-like region of the $\omega,~k$ plane above some $k$. 
 For the dispersion relation (\ref{JRDispersion2}) this is the case for $k>\omega_p/\sqrt(1-u^2)$. This is different for the high-temperature plasma, where longitudinal and transverse plasma modes only appear in the time-like region, $|x|=|\omega/k|>1$.  In the quantum liquid scenario one can expect that the modes with low phase velocity $|x|<u$ suffer severe Landau damping because they accelerate the slower moving charges and decelerate those moving faster than the wave \cite{Weldon, Ichimaru}. A charge moving with a velocity that is lower than the speed of plasmon propagation can only excite these modes and not the modes with intermediate phase velocities $u<|x|<1$, which are undamped \cite{Ichimaru, Weldon}. The qualitative properties of the color wake can in this case expected to be analogous to those of the high temperature plasma, namely that the charge carries only a screening color cloud with it and Mach cones do not appear. 

If the colored parton travels with a velocity $v>u$ that is higher than the speed of sound in the medium, modes with an intermediate phase velocity $u<|x|<1$ can be excited. The emission of these plasma oscillations induced by supersonically traveling particles is analogous to Cherenkov radiation. This can be expected to lead to the emergence of Mach cones in the induced charge density cloud with the opening angle
\bea \label{JRMach}
\Delta \Phi={\rm arccos}\left(u/v\right).
\eea

The effect is well known in solid state physics \cite{Ichimaru} and is analogous to the Mach cones induced by fast heavy ions in electron plasmas \cite{Schaefer1, Schaefer2, Groeneveld}. The physics of shock waves in relativistic heavy-ion collisions has also been discussed in \cite{Glassgold, Scheid:1973, Baumgardt:1975qv}.

To be specific the following dielectric function is assumed in accordance with (\ref{JRDispersion2}):
\bea \label{JRNonBloch}
\epsilon_L=1+\frac{\omega_p^2/2}{u^2 k^2 - \omega^2 + \omega_p^2/2}\,\, \, \, (k\le k_c)  \, \,.
\eea Note that this differs from the classical, hydrodynamical 
dielectric function of Bloch \cite{Bloch}, since the latter one 
is singular at small $k$ and $\omega$ due to phonon contributions 
which cannot mix with colored plasmons.

This dielectric function (\ref{JRNonBloch}) has the plasmon mode (see Eqn.~\ref{JRDispersion2}) which extends into the space-like region of the $\omega, k$ plane, see Fig.~\ref{JRfigure0}. The qualitative induced wake structure in such a quantum liquid scenario is general, namely a Mach shock wave structure for a `supersonically' traveling color source as discussed above. In that sense the principal findings can be expected to hold generally for a quantum liquid with a plasmon branch similiar to (\ref{JRNonBloch}) independent of the exact form of the dielectric function.

To be specific a speed of plasmon propagation of $u/c= 1/\sqrt{3}$ is assumed (compared that to $u/c=\sqrt{3/5}$ in the HTL approximation and to $u/c \approx \sqrt{0.2}$ \cite{Shuryak2, Venugopalan} for a hadron resonance gas). 

\begin{figure}
 \par\resizebox*{!}{0.20\textheight}{\includegraphics{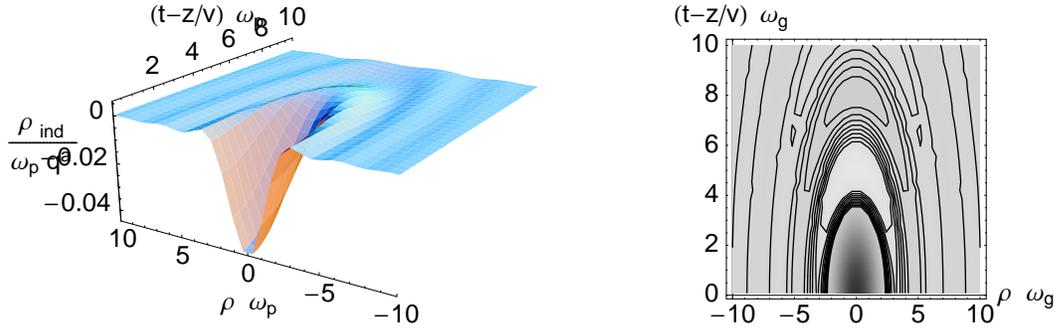}} \par{}
\caption{Spatial distribution of a induced charge density from a  jet with 
high momentum and a fixed color charge $q^a$ that is traveling with $v=0.55c<u$. The right plot shows equi-charge lines in the density distribution. The density profile is that of a cloud traveling with the color charged jet. 
\label{JRfigure2}}
\end{figure}

\begin{figure}
 \par\resizebox*{!}{0.25\textheight}{\includegraphics{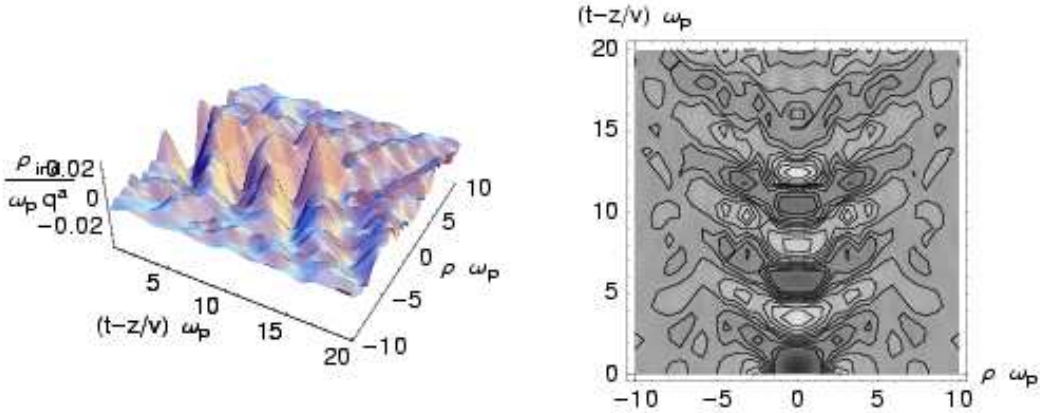}} \par{}
\caption{Spatial distribution of the induced charge density from a  jet with high momentum and fixed color charge $q^a$ that is traveling with 
$v=0.99c>u=\sqrt{1/3}c$. The right plot shows equi-charge lines in the density distribution. 
\label{JRfigure31}}
\end{figure}

Figures \ref{JRfigure31}a,b show the induced charge density for a colored particle traveling `supersonically' at $v=0.99 c>u$ which is calculated using eq.~(\ref{JRcharge2}). The integration area has been restricted to the region $k<k_c=2 \omega_p$\footnote{Note that the expressions in \cite{Schaefer1, Schaefer2} do not correspond to $k<k_c$, but $\kappa<k_c$.} . Numerical consistency has been checked also via calculation of the induced current densities and the continuity equation. On the other hand a `subsonically' traveling particle with 
$v=0.55c<v$ induces a charge density profile which is qualitativly analogous
to the high-temperature plasma, namely a comoving screening cloud, see Fig.~\ref{JRfigure2}.

\section{Observable consequences}
\label{JRobservable}

It can be expected that the phenomenon of Mach cones in the charge density translates into Mach cones in the particle density and should eventually lead 
to a directed emission of secondary partons from the plasma. This effect
has an analogoy in solid state physics where Mach cones induced by fast
heavy ions in an electron plasma lead to an emission of electrons that
have been carried within the wake \cite{Schaefer1,Schaefer2}. This effect
has been studied experimentally \cite{Groeneveld}.

If that scenario of a strongly coupled QCD plasma is realized in the matter created in an ultrarelativistic heavy ion collision, one could expect to observe these cones in the actual distribution of secondary particles associated with jets at RHIC, a signature also proposed in \cite{Stoecker, Shuryak}. 

Indeed preliminary data from the PHENIX and STAR experiments (see e.g. Fig.~1 in \cite{STAR} and preliminary data from PHENIX) show such effects in the background distribution of secondary particles in the azimuthal angle $\Delta \phi$. 
The peak near zero degree corresponds to secondaries from the outmoving jet, 
particles from the companion jet result in a distribution with a clear maximum near $\Delta \phi=\pi$ for pp collisions, where no medium effects are present. In Au-Au collisions there is a distribution with two maxima around the $\Delta \phi=\pi$ position. These are located at $\Delta \phi \approx \pi  \pm 1.1$.  One can argue \cite{Shuryak} that such an effect could possibly be traced back to a Mach shock front (predicted in a hydrodynamical framework) traveling with the side-away jet  leading to maxima in the distribution at about  $\Delta \Phi = \pi \pm {\rm arccos} (u/v)$.

Given the confirmation of this effect in the data mentioned, this would clearly indicate that the first scenario, viz. a pQCD plasma, is not realized in RHIC experiments. A further decision on the possible occurence of 
Mach cone structures could be deduced by studying correlations in the secondaries as proposed by the PHENIX collaboration. Indication of such a Mach cone structure could reveal properties of the plasma and it's plasmons in general, which is a fascinating perspective.  
The speed of plasma propagation could be also determined. In fact if the maxima at $\Delta \phi \approx \pi \pm 1.1$ are experimentally confirmed, it would correspond to $u/c\approx\sqrt{0.2}$ \cite{Shuryak}. It is interesting to note, that a study of  the angular structure of the collisional energy losses of a hard jet in the medium would also support such an observation: the incident hard jet's scattering angle vanishes in the relativistic limit - leading to the jet's propagation along a straight line - whereas the expectation value of the scattering angle $\Theta$ of a struck "thermal" particle is $\left<\Theta\right>\approx 1.04$ \cite{Lokhtin} which is close to $1.1$.

\section{Conclusions}
\label{JRconclusions}

The properties of the charge density wake of a colored hard partonic jet traveling  through a QGP plasma in the framework of linear response theory has been discussed. Two different scenarios have been studied, namely a high temperature QGP at $T \gg T_c$ described in the HTL approximation, and a description of a strongly coupled QGP  (sQGP) behaving as a quantum liquid. 
It was found that the structure of the wake corresponds to a screening color cloud traveling with the particle in the case of the high temperature plasma and in the case of a quantum fluid, if the velocity of sound in the plasma is not exceeded by the jet in the latter case. The structure of the wake is changed considerably in comparison to the former cases, if the jet's velocity exceeds the plasma's speed of sound and the collective modes have a dispersion relation extending in the space-like region. Then the induced parton density exhibits the characteristics of Mach waves trailing the jet at the Mach angle.

It is argued that this effect could be used to constrain theoretically possible scenarios by experimental analysis via the measurments of angular distribution of the secondary particle cones of jet events in RHIC. First indications of the observation of a Mach shock phenomenon in RHIC in the quenched jet's secondary particle distribution from PHENIX and STAR data were discussed. 

In general secondary particle distributions can be used to provide methods of probing the QCD plasma's collective excitations experimentally.
\bigskip

\section*{Acknowledgements}
The results presented here rely on 
work \cite{Wake} done in a collaboration 
with Berndt M\"uller whom I want to thank in the first place.  
 I thank Purnendu Chakraborty and Munshi G. Mustafa for drawing my attention to a numerical inconsistency regarding an earlier version of Fig.~\ref{JRfigure1}. I also want to thank Steffen Bass, Jorge Casalderry,  Rainer Fries, Roy Lacey, Abhijit Majumder, and Wolf Holzmann for interesting discussions.
This work was supported in part by U.~S.~Department of Energy under grants DE-FG02-96ER40945 and DE-FG02-05ER41367. I thank the Alexander von Humboldt Foundation for support as a Feodor Lynen Fellow. 

\bibliography{u4}

\end{document}